\documentclass{aa}  

\usepackage{xcolor}
\usepackage{graphicx}
\usepackage{txfonts}
\usepackage{float}
\usepackage{amsmath}

\newcommand\Rey{\mbox{Re }}  
\newcommand\Reym{\mbox{Rm }}  

\newcommand\Bo{\mathbf{\overline{B}}_0}

\makeatletter
\begin{document}

   \title{Dissipative effects on the sustainment of a magnetorotational dynamo in Keplerian shear flow}
	
   \author{A. Riols\inst{1,2}\and F. Rincon\inst{1,2}
 		  \and C. Cossu \inst{3} \and G. Lesur \inst{4,5} \and G. I. Ogilvie \inst{6} \and P. -Y. Longaretti \inst{4,5}
          }	
   \institute{
   			  Universit\'e de Toulouse; UPS-OMP; IRAP: Toulouse, France
            \and
             CNRS; IRAP; 14 avenue Edouard Belin, F-31400 Toulouse, France
           \and            
            CNRS; Institut de M\'ecanique des Fluides de Toulouse (IMFT), Allée du Professeur Camille Soula, 31400 Toulouse, France
            \and            
            Univ. Grenoble Alpes, IPAG, F-38000 Grenoble, France
	    \and
	    CNRS, IPAG, F-38000 Grenoble, France
             \and            
            Department of Applied Mathematics and Theoretical Physics, University of Cambridge, Centre for Mathematical Sciences, Wilberforce Road, Cambridge CB3 0WA, United Kingdom         
            }	

	\date{\today}
   \abstract
{The magnetorotational (MRI) dynamo has long been considered one of the possible
drivers of turbulent angular momentum transport in astrophysical accretion disks. 
However, various numerical results suggest that this dynamo may be
difficult to excite in the astrophysically relevant regime of
magnetic Prandtl number ($\text{Pm}$) significantly smaller than unity, for
reasons currently not well
understood. The aim of this article is to present the first results of
an ongoing numerical investigation of  the role of both linear and
nonlinear dissipative effects in this problem. Combining a 
parametric exploration and an energy analysis of incompressible 
nonlinear MRI dynamo cycles representative of the transitional
dynamics in large aspect ratio shearing boxes, we
find that turbulent magnetic diffusion makes the excitation and
sustainment of this dynamo at moderate magnetic Reynolds number
($\text{Rm}$) increasingly difficult for decreasing $\text{Pm}$. This
results in an increase in the critical $\text{Rm}$ of the dynamo for
increasing kinematic Reynolds number ($\text{Re}$), in agreement with
earlier numerical results. Given its very generic nature, we argue
that turbulent magnetic diffusion could be an important determinant of
MRI dynamo excitation in disks, and may also limit the efficiency of
angular momentum transport by MRI turbulence in low $\text{Pm}$ regimes.}


 \titlerunning{Dissipative effects on the sustainment of a magnetorotational dynamo in Keplerian shear flow}
   \maketitle

%

\section{Introduction}
   Magnetorotational instability (MRI) occurs in differentially
   rotating flows whose angular velocity decreases with distance
   to the rotation axis \citep{velikhov59,chandra60,balbus91}
   and is the most commonly invoked excitation mechanism of angular
   momentum-transporting turbulence in accretion disks
   \citep{balbus98}.  In a uniform
   magnetic field ${\mathbf{B}}$, the MRI amplifies arbitrarily small
   perturbations exponentially  and breaks down nonlinearly into MHD
   turbulence \citep[e.g.][]{hawley95}. The transport efficiency of MRI
   turbulence continues to be debated, and may be reduced in
   the astrophysically relevant regime of low magnetic Prandtl number
   Pm, the ratio between kinematic viscosity and magnetic diffusivity
   \citep{lesur07,balbus08}. 

   Another question in this context is that of the origin of the
   MRI-supporting magnetic field. In some cases, this field may be 
   generated by an internal disk dynamo process which could bootstrap MHD
   turbulence in the disk independently of its magnetic environment 
   \citep{balbus98, donati05}.
   Early simulations by \citet{hawley96} in the so-called
   zero net flux configuration appropriate to this problem 
   showed that such a dynamo is indeed possible and is intimately
   coupled to the MRI (see also \citet{branden95}), but its
   viability in disks has since been questioned by numerical studies 
   suggesting that it may be impossible to excite at low Pm \citep{fromang07b}, 
   although the physical reasons for this are not yet clear \citep{bodo11,kapyla11,oishi11,simon11}.  

The aim of this article is to seek a physical explanation for this
behaviour by exploiting recently discovered dynamical properties
of this subcritical dynamo mechanism \citep{rincon07b,rincon08,lesur08},
whose principles are otherwise rather simple:
starting from a zero net-flux axisymmetric weak poloidal field, a
larger toroidal field is generated through the $\Omega$
effect. This field is MRI-unstable to non-axisymmetric MHD
perturbations, whose growth results in a nonlinear
   electromotive force (EMF) that sustains (and can also reverse) the
   axisymmetric field. Recent work suggests that three-dimensional cyclic nonlinear solutions provide the first germs of excitation of the 
   dynamo in shearing box simulations \citep{Herault2011, riols2013} 
   and possibly form the backbone of the ensuing self-sustained MHD 
turbulence.  Parametric studies of cycles representative of the
transitional dynamics, complemented with an analysis of their
energetics, may therefore prove useful to understand how dissipative 
effects affect the dynamo transition as a whole. 
Here, we present the first results of an ongoing
numerical investigation of this kind. We focus on the simpler
case of incompressible dynamics in large aspect ratio shearing boxes,  
which includes most of the fundamental physical complexity of the
problem, except for stratification and boundary
effects. An exhaustive study of different configurations
will be presented in a subsequent paper. \\

{The equations and numerical framework used in this article are presented in Sect.~2. In Sect.~3, we study the characteristics of the transition in elongated shearing boxes using generic incompressible numerical simulations, in order to check if the results of \citet{fromang07b} on the Pm-dependence of the transition extend to such configurations. We also investigate whether cycles still provide the first germs of MRI dynamo chaos at kinematic Reynolds number Re larger than studied by \citet{riols2013}. In Sect.~4, we compute the existence boundaries of several cycles in the magnetic versus kinematic Reynolds number parameter plane and analyse their energy budget to identify physical effects affecting the dynamics in the vicinity of $\text{Pm}\sim 1$. Additional numerical experiments aiming at investigating the conditions of excitation of the dynamo, and why it appears harder to excite at low Pm, are presented in Sect. 5. A short discussion concludes the paper.}

\section{Equations and numerical framework\label{model}}
\subsection{{Model}}

The equations and numerical framework are the same as in the work of \citet{Herault2011} and \cite{riols2013} and have already been described in detail in
these papers. We use the cartesian local shearing sheet description of differentially rotating flows \citep{goldreich65}, whereby the axisymmetric differential rotation is approximated locally by a linear shear flow  $\mathbf{U}_x=-Sx\,\mathbf{e}_y$, and a uniform rotation rate $\mathbf{\Omega}=\Omega\,\mathbf{e}_z$, with  $\Omega=(2/3) S$ for a Keplerian equilibrium. Here $(x,y,z)$ are respectively the shearwise, streamwise and spanwise directions, corresponding to the radial, azimuthal and vertical directions in accretion disks. {We refer to the $(x,z)$ projection of vector fields as their poloidal component and to the $y$ direction as their toroidal component.} {Stratification and compressibility effects are ignored for simplicity.} The evolution of the three-dimensional velocity {field} perturbations $\mathbf{u}$ and magnetic field $\mathbf{B}$ is governed by the three-dimensional incompressible, dissipative MHD equations:
{\begin{equation}
\frac{\partial{\mathbf{u}}}{\partial{t}}-Sx\frac{\partial{\mathbf{u}}}{\partial{y}}+\mathbf{u}\cdot\mathbf{\nabla u} = -2\mathbf{\Omega}\times\mathbf{u}+Su_x\mathbf{e}_y-\mathbf{\nabla}\Pi+\mathbf{B}\cdot\mathbf{\nabla B}+\nu\mathbf{\Delta u},
\label{velocity_eq}
\end{equation} 
\begin{equation}
\frac{\partial{\mathbf{B}}}{\partial{t}}-Sx\frac{\partial{\mathbf{B}}}{\partial{y}}  = -SB_x\mathbf{e}_y+\nabla\times(\mathbf{u}\times\mathbf{B})+\eta\mathbf{\Delta B},
\label{magnetic_eq}
\end{equation}
\begin{equation}
\nabla \cdot\mathbf{u}=0, \quad \nabla\cdot\mathbf{B}=0.
\label{div}
\end{equation}}The kinematic and magnetic Reynolds numbers are defined by $\text{Re}=SL^2/\nu$ and
$\text{Rm}=SL^2/\eta$, where $\nu$ and $\eta$ are the constant kinematic viscosity and magnetic diffusivity, $L$ is a typical scale of the
spatial domain and time is measured with respect to
$S^{-1}$. The magnetic Prandtl number is 
$\text{Pm}=\nu/\eta=\text{Rm}/\text{Re}$. {$\Pi$ is the total of fluid plus magnetic pressure divided by the uniform density.} $\mathbf{B}$ is expressed in
terms of an alfv\'enic velocity. Both $\mathbf{u}$ and $\mathbf{B}$ are
measured in units of $SL$. 

{\subsection{Numerical methods}}

We use the SNOOPY code \citep{lesur07} to perform direct numerical simulations (DNS) of Eqs. \eqref{velocity_eq}-\eqref{div}. This code provides a spectral implementation of the so-called numerical shearing box model of the shearing sheet, in a finite domain {of size $(L_x,L_y,L_z)$, at numerical resolution $(N_x,N_y,N_z)$}. The $x$ and $y$ directions are taken as periodic while shear-periodicity is imposed in $x$. 
{A discrete spectral basis
of shearing waves with constant $k_y$ and $k_z$ wavenumbers and
constant shearwise Lagrangian wavenumber $k_{x_0}$ is used to represent
the fields in the sheared Lagrangian frame. The shearing of nonaxisymmetric  perturbations in this model ($k_y\neq0$) is described using time-dependent Eulerian shearwise wavenumbers, $k_x(t)=k_{x_0}+Sk_yt$.} {Shearing waves are "leading" when $k_xk_y<0$ and "trailing" when $k_xk_y>0$.}\\

Nonlinear periodic solutions are computed with the Newton-Krylov solver PEANUTS interfaced to SNOOPY, and followed in parameter space using arclength continuation. Almost all the results presented in the paper are for a maximum resolution of $(48,48,72)$, ensuring convergence for all parameters considered, {except for some of the results of Sect.~\ref{Pm_dependence} which required a higher resolution.}

{\subsection{Symmetries and aspect ratio choice}}
\label{aspect_ratio}

\begin{figure}
 \centerline{\includegraphics[width=\columnwidth]{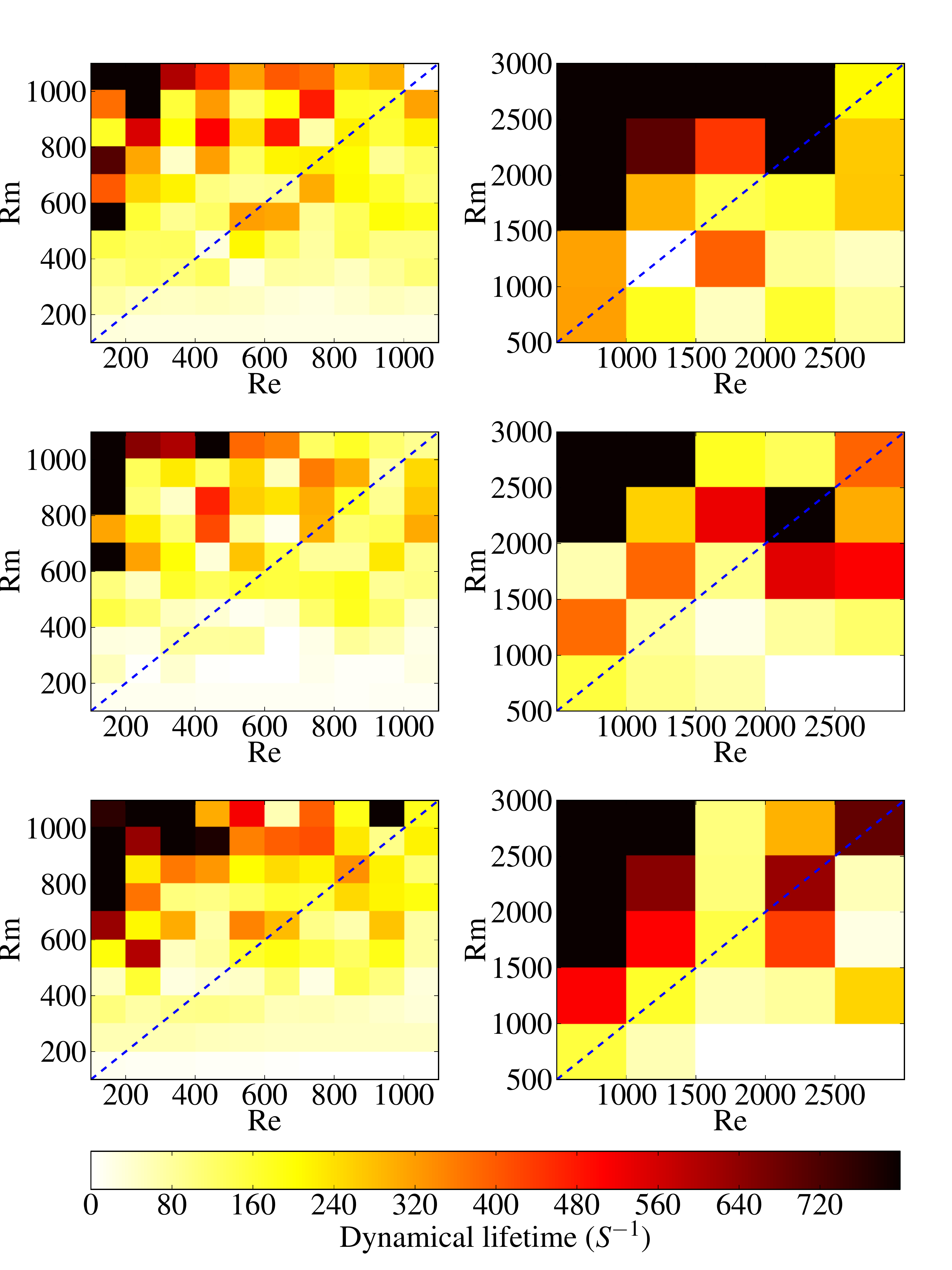}}
 \caption{{Maps of the dynamical lifetime in generic DNS, as a function of Re and Rm ($L_x=0.7$, $L_y=20$, $L_z=2$). Each row is for a different random initial condition. The second and third maps (from top to bottom) use the same noise realization (see text) but different amplitudes ($A$=5 for the second case and $A$=2.5 for the third case). The maps on the left are for simulations at} {mild Re and Rm that can be conducted at moderate} {numerical resolution ($48 \times 48 \times 72$). The maps on the right are} {for higher Re and Rm,  requiring} {a higher resolution ($96\times 96 \times 128$). The dashed line corresponds to Pm=1.}}
\label{map_turb_Pm}
\end{figure}

\begin{figure*}
\centering
\includegraphics[width=0.96\textwidth]{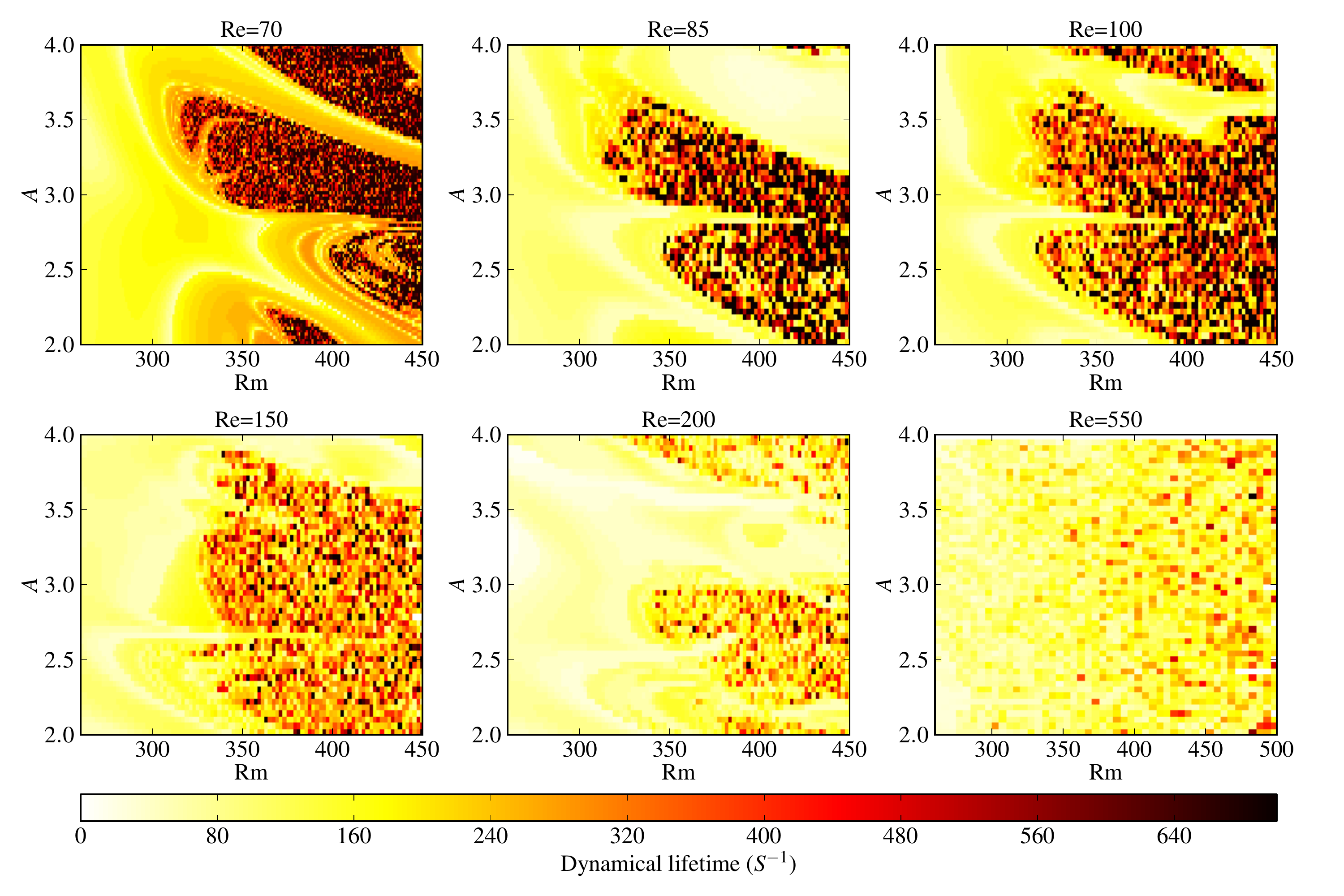}
\caption{Maps of dynamical lifetimes as a function of
  \text{Rm} and initial perturbation amplitude $A$, for a fixed noise
  realization {(see text)} and six different Re. 
   {The resolution is ($\delta
   A=0.04,\delta \text{Rm}=2$) except for $\text{Re}=70$ where ($\delta A=0.02,\delta \text{Rm}=1$) and for $\text{Re}=550$ where $\delta\text{Rm}=5$.}}
\label{map_turb_ARm}
\end{figure*}
{The dynamics in the} {transitional} {regimes typical of simulations displaying recurrent dynamics  (Re and Rm of a few} {hundreds to} {thousands) is already quite complex and clearly involves} {a large number of  cyclic solutions.}  {Given the current state of understanding of the dependence of the MRI dynamo transition on dissipative effects, our strategy to make progress on this problem is to simplify the dynamics as much as possible and focus on a few simple dynamo cycles which encapsulate the basic physics of the dynamo. To achieve this}, we enforce that the dynamics takes place in a symmetric subspace to facilitate the analysis (this does not compromise the underlying dynamical complexity, {see Sect.~3 of \citet{riols2013}}), and notably monitor the axisymmetric MRI-supporting field $\overline{\mathbf{B}}$ ($y$-average of $\mathbf{B}$), more specifically its energetically dominant Fourier mode $\Bo(z,t)=\Bo(t)\cos(k_{z_0}z)$ with $k_{z_0}=2\pi/L_z$. {We also restrict our study to the large aspect ratio configuration $(L_x,L_y,L_z)=(0.7,20,2)$ already employed by \citet{Herault2011} and \citet{riols2013}, and Re and Rm values in the range of a few hundred, which effectively guarantees that only a small number of simple nonlinear cyclic solutions are excited (see Sect.~III.~D of \citet{Herault2011} for a detailed explanation). Different aspect ratio configurations will be explored in a future paper.}

{\section{Numerical exploration of the dynamo transition}}
{\subsection{Dynamical lifetime of DNS in the (Re,Rm) plane}
\label{Pm_dependence}
As mentioned in the introduction, the numerical study of \citet{fromang07b}, for a single box size $(\pi,2\pi,\pi)$, in a compressible isothermal case, suggests that zero net flux MRI turbulence cannot be sustained for Pm below some critical $\rm{Pm}_c$. The transitional dynamics in this box {involves very intricate nonlinear interactions between many shearing waves and is therefore particularly difficult to analyse. As explained previously, we found it more convenient here to use an elongated box in which fewer shearing waves are active to address the problem of the nature of the Pm-dependence of the transition.\\}}

{As a preliminary step, we first ensured that} {the results obtained by \citet{fromang07b} pertain to our large aspect ratio configuration. For this purpose, we used a cartography procedure similar to that described in \citet{riols2013}. We performed a series of DNS for different \Rey and Rm, using the same random initial condition. The latter was generated as follows: for each field component, we generated a set of random complex Fourier modes and normalized the total energy density to obtain a particular {white noise incompressible} {"realization". For $\mathbf{u}$ and $\mathbf{B}$, a given zero net-flux initial condition is obtained} {by multiplying this particular} noise realization by an amplitude factor $A$} {(see \cite{riols2013} for details).} {The typical dynamical lifetime measured in each DNS was then plotted on a two dimensional map covering the (Re,Rm) grid. 
To check whether the results were generic, we performed the same experiment for three different initial conditions. The first and second ones were constructed from different noise realizations but the same amplitude $A=5$. The third one used the same noise realization as the second one but with $A=2.5$ (shooting along the same direction in phase space but at different distances from the laminar state).} \\

{Figure~\ref{map_turb_Pm} (left) shows the corresponding dynamical lifetime maps for \Rey and \Reym between 100 and 1000, computed for a numerical resolution $48\times 48 \times 72$. The maps on the right are for the same initial conditions, but extend to higher \Rey and \Reym (from 500 to 3000). They required a larger numerical resolution ($128\times 128\times 96$). All the DNS whose dynamical lifetime exceeds 600 $S^{-1}$ are systematically on or above the Pm $\sim$ 1 line.  At low Re, the dynamics seems to be sustained only for \Reym larger than some critical $\rm{Rm}_c$. At higher Re, the transition border visually follows a Pm $\simeq\rm{Pm}_c$ line, with $\rm{Pm}_c$ of the order of unity. This behaviour suggests that the transition border is indeed  similar to that obtained by \citet{fromang07b}, and does not depend on the shearing box aspect ratio or compressibility of the fluid}{, at least on the qualitative level.}
\subsection{Transition maps}

To justify our interest in cycles, {we then attempted to check whether}
the conclusion of \cite{riols2013}, that chaotic dynamo action
at $\text{Re}=70$ results from their global bifurcations,
extends to larger $\text{Re}$. The simplest signature of this effect
is in the form of fractal-like sets of initial perturbations for which the
dynamics is long-lived. To check this, we performed several series of
DNS spanning a range of $\text{Rm}$ and initial conditions, each of them
generated from a unique noise realization {(using the same procedure as in Sect.~\ref{Pm_dependence}) and varying the perturbation amplitude $A$.} We constructed maps of the dynamical lifetimes for each run as a function
of $\text{Rm}$ and $A$, for $\text{Re}=(70, 85, 100, 150, 200, 550)$.

Figure~\ref{map_turb_ARm} shows that the boundary separating the
regions in phase space where the dynamics is long-lived from those
where perturbations decay rapidly has the same qualitative
fractal-like structure for all $\text{Re}$. Long-lived simulations are
characterized by recurrent dynamics reminiscent of nonlinear cycles,
suggesting that the excitation of the MRI dynamo is indeed tied to that
of cycles.

{At the largest Re considered though, the fractal-like features appear  to be smoothed out, and the correspondence between recurrent dynamics and chaotic  flows less pronounced. We also note that at  $\text{Re}=550$, the transition  border seems to be at higher Rm than at  $\text{Re}=70$, in line with the results of \cite{fromang07b} and} {the results of} { Sect.~\ref{Pm_dependence}.} \\

{Overall, the previous results suggest that the excitation of
self-sustaining MHD turbulence in this problem is related to the existence of MRI dynamo cycles and their global bifurcations. This vindicates the idea that studying how the dynamics of simple nonlinear cycles is affected by changes in Re and Rm may be useful to identify the physical mechanisms responsible for the 
Pm-dependence of the transition.}

\section{{Parametric study of MRI dynamo cycles}}
\subsection{Existence boundaries in the (Re,Rm) plane}
\label{boundary}

\begin{figure}
\centering
\includegraphics[width=1.\columnwidth]{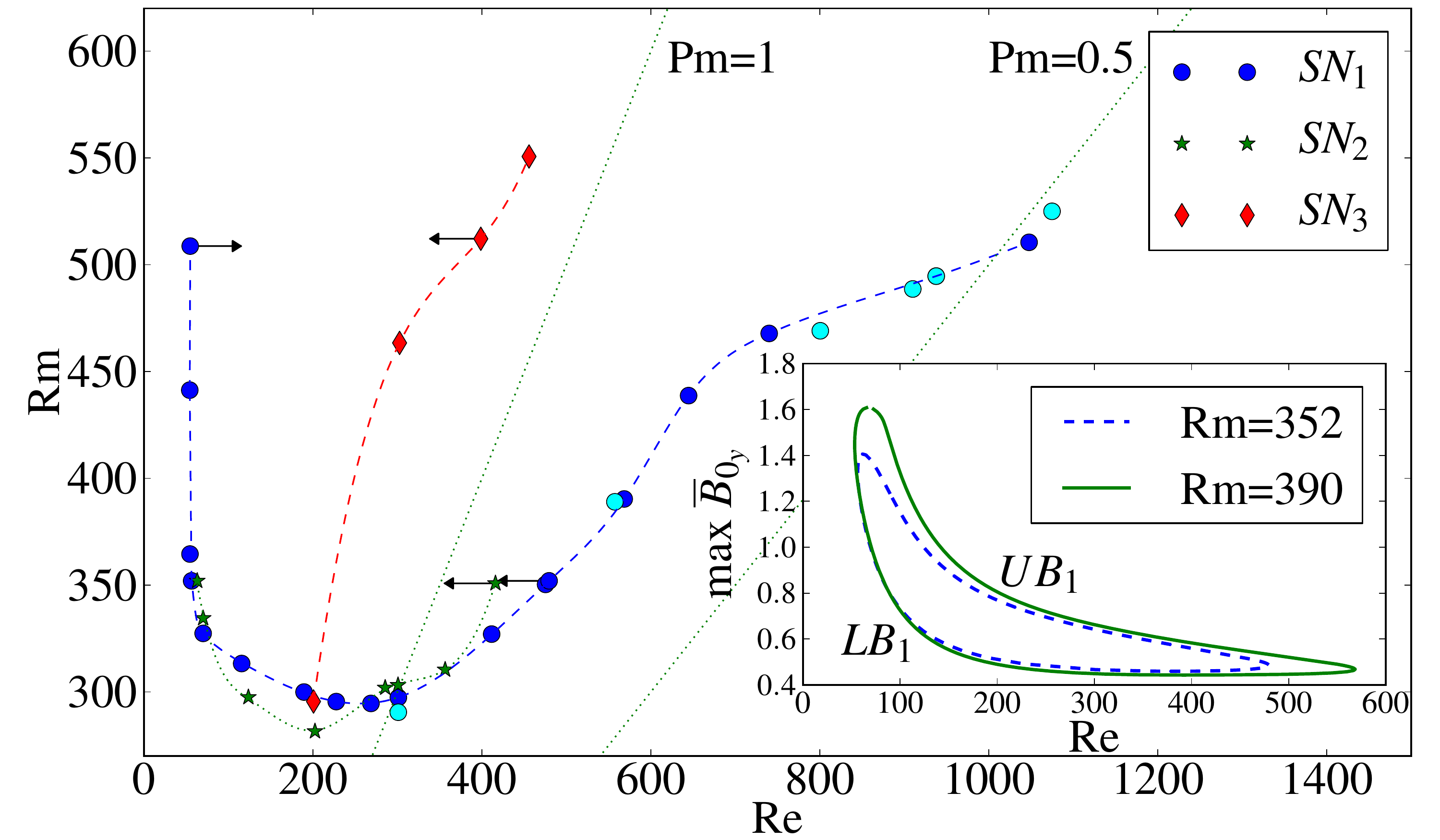}
 \caption{Existence boundaries of three cycle pairs
   (dashed lines) in the ($\text{Re}$,$\text{Rm}$) plane and critical
   $\text{Re}_c(\text{Rm})$, $\text{Rm}_c(\text{Re})$ of saddle
   nodes obtained by continuation (symbols). Black arrows
   indicate the regions in which the cycles exist. {For $SN_1$, dark blue/black bullets correspond to low resolution results $(24\times12\times36)$, and light blue/grey bullets} {are for double resolution} {$(48\times 24\times 72)$}. Inset: selected continuation curves of $SN_1$ at fixed Rm.}
\label{fig_frontiere}
 \end{figure}
Motivated by the previous results, we investigated the domains of existence in parameter space 
of three pairs of cycles $SN_1$, $SN_2$ and $SN_3$ born out of saddle
node bifurcations at pair-specific critical $\text{Rm}_c(\text{Re})$.
$SN_1$ and $SN_2$ have already been documented by \citet{riols2013}
(see their Figs. 7-8), while $SN_3$ was found more recently.

Continuation with respect to $\text{Re}$ at fixed $\text{Rm}$
of the lower and upper branches $LB_1$ and $UB_1$ of $SN_1$ 
(Fig.~\ref{fig_frontiere}, inset) shows that they only exist in a
finite range of $\text{Re}$, whose extent widens as $\text{Rm}$
increases. $SN_2$ and $SN_3$ behave similarly (not shown). 
The existence boundaries of all cycle pairs in the
($\text{Re}$,$\text{Rm}$) plane were constructed by
combining all critical $\text{Rm}_c(\text{Re})$ and $\text{Re}_c(\text{Rm})$
obtained by continuation (Fig.~\ref{fig_frontiere}). 
For $\text{Rm}$ in the $300-500$ range, all cycles disappear at low
enough $\text{Pm}$ and their $\text{Rm}_c$ increases with
$\text{Re}$. This behaviour, reminiscent of the results of
\cite{fromang07b}, is investigated below (the seemingly large 
differences between their transitional $\text{Re}$, $\text{Rm}$ and 
ours are essentially due to aspect ratio differences and do not
reflect fundamental physical differences). {For $SN_1$, we} {plotted the critical points obtained for both low resolution simulations $(24\times 12\times 36)$ and at higher resolution $(48\times 24\times 72)$.  The boundary appears to be almost independent of resolution.} Note that we found it very difficult to explore the strongly nonlinear regime $\text{Rm}>500$ {at the maximum resolution considered}. Tentative results at this resolution (not shown) suggest that $SN_1$ may have an upper boundary in $\text{Rm}$ in this regime.

\subsection{Magnetic energy budgets of MRI dynamo cycles}
As the MRI dynamo rests on the sustainment of the axisymmetric MRI-supporting field $\Bo$ against {dissipative processes} through nonlinear interactions of non-axisymmetric modes, analysing the magnetic energy
budget of dynamo cycles, most importantly transitional lower branch
saddles such as $LB_1$, may give useful physical insights into the
$\text{Pm}\leq 1$ regime. To do this, we write
\begin{equation}
\mathbf{B}=\Bo + \mathbf{b}_1+ \sum_{j\geqslant 2} \mathbf{b}_j  \quad \text{and} \quad 
\mathbf{u}= \mathbf{u}_1+ \sum_{j\geqslant 2}\mathbf{u}_j~,
\end{equation}
where $\mathbf{u}_1$ and $\mathbf{b}_1$ stand for non-axisymmetric
``MRI wave'' perturbations supported by shearing waves with
wavenumbers $\vert k_y\vert =k_{y_0}=2\pi/L_y$, $k_x(t)=Sk_y\,t$,
$\vert k_z\vert =k_{z_0}$ for $\mathbf{u}_1$ and $k_z=0$ for
$\mathbf{b}_1$. 

The sum over $j\geq 2$ stands for all other (smaller-scale)
fluctuations. 

\begin{figure}
\centering
\includegraphics[width=1.\columnwidth]{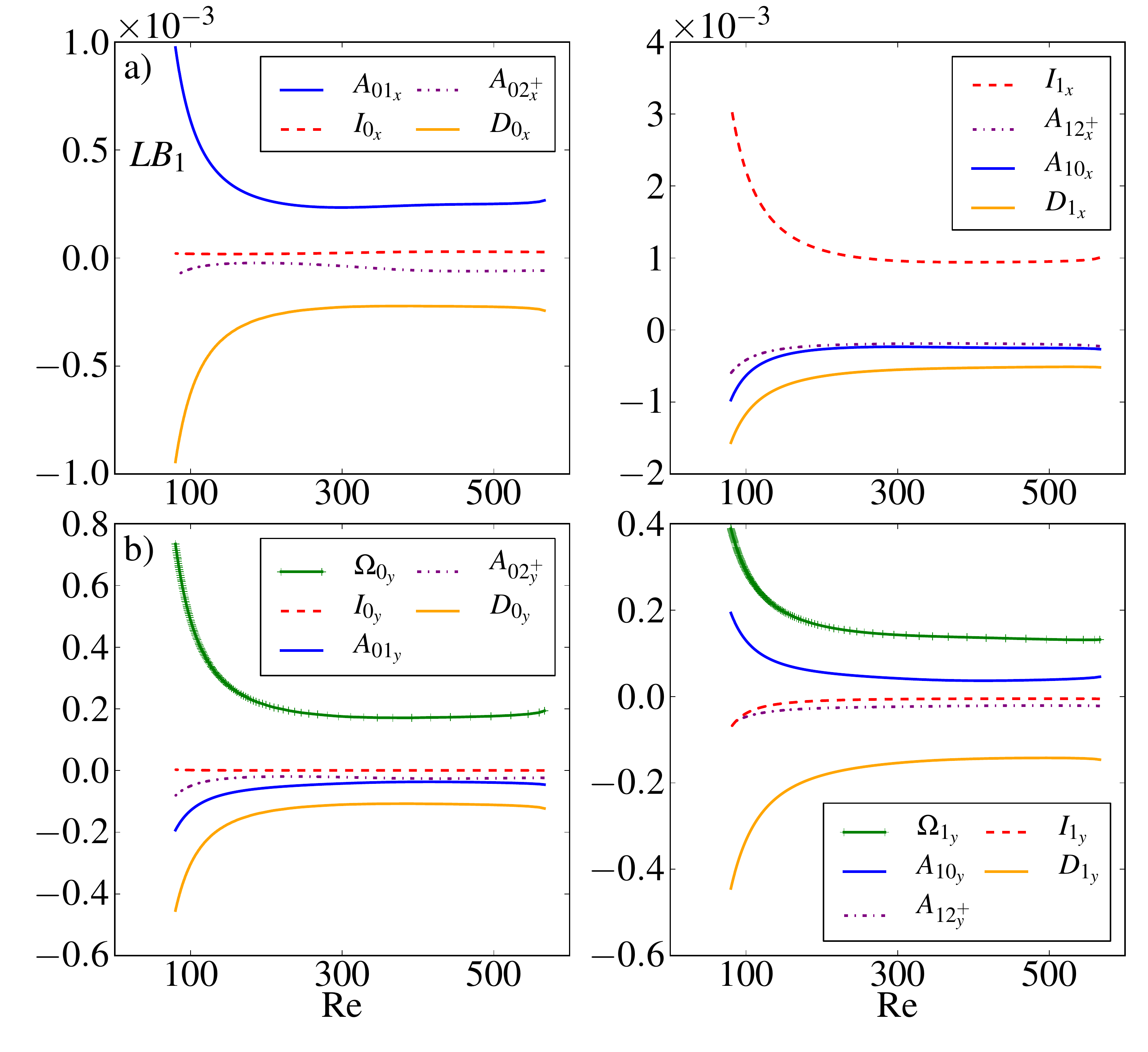}
 \caption{Magnetic energy budgets as a
   function of $\text{Re}$. a) $x$ projection of the budgets
   for $\Bo$ (left) and $\mathbf{b}_1$ (right) for $LB_1$ at
   $\text{Rm}=390$. b) corresponding $y$ projection. }
\label{fig_c1Re}
 \end{figure}

We first integrate the energy equation for $\Bo(z,t)$ over the volume
and half cycle period $T/2=S^{-1}L_y/L_x$. Denoting this operation by
$\langle\rangle$ and taking into account that cyclic magnetic reversals
change $\Bo$ into $-\Bo$, we obtain the energy budget for $\Bo$,
\begin{equation}
\label{B0budget}
{\mathbf{\Omega}}_0+{\mathbf{I}}_0+{\mathbf{A}}_0+{\mathbf{D}}_0=\bf{0}, \quad  \text{where} \\
\end{equation}
\begin{equation}
{\mathbf{\Omega}}_0= -S\langle\overline{B}_{0_y}\overline{B}_{0_x}\rangle\,\mathbf{e}_y, \quad  \quad
{\mathbf{I}}_0= \langle \Bo  \circ \overline{\mathbf{B}\cdot \mathbf{\nabla}\mathbf{u}} \rangle,
\end{equation}
\begin{equation}
 {\mathbf{D}}_0=-\eta\,{k_{z_0}^2}\langle  \Bo \circ \Bo  \rangle,  \quad  \quad
 {\mathbf{A}}_0={\mathbf{A}}_{01}+{\mathbf{A}}_{02^+},\\
 \end{equation}
 \begin{equation}
{\mathbf{A}}_{01}=-\langle \Bo \circ  \overline{\mathbf{u}_1\cdot \mathbf{\nabla}\,\mathbf{b}_1} \rangle, \quad   \quad
{\mathbf{A}}_{02^+}=-\langle \Bo \circ  \overline{\mathbf{u}\cdot \mathbf{\nabla}\,\mathbf{B}} \rangle -{\mathbf{A}}_{01},
\end{equation}
and $\circ$ is the Hadamard (entrywise) product; ${\mathbf{\Omega}}_0$ is
the energy provided by the linear stretching of $\overline{B}_{0_x}$
by the shear ($\Omega$ effect) and ${\mathbf{I}}_0$ is a nonlinear
induction term; ${\mathbf{A}}_0$ is the
magnetic energy exchanged with other modes through nonlinear advection
and ${\mathbf{D}}_0$ is the ohmic dissipation; ${\mathbf{A}}_{01}$ is
the energy exchanged with the MRI-unstable waves and
${\mathbf{A}}_{02^+}$ is the energy exchanged with all $j\geq2$
modes. Figure~\ref{fig_c1Re}a-b) (left) display the $x$ and $y$
projections of Eq.~(\ref{B0budget}) for $LB_1$ as a function of $\text{Re}$, 
at fixed $\text{Rm}$ ($UB_1$ is more energetic but behaves similarly).
The MRI-supporting azimuthal field $\overline{B}_{0_y}$ loses energy
through laminar dissipation ${D}_{0_y}$, but also through a nonlinear advective
transfer to other modes ${A}_{0_y}<0$, which therefore acts
as a weakly nonlinear (``turbulent'') diffusion.
The $\Omega$ effect is the only net source term for $\overline{B}_{0_y}$,
therefore the sustainment of $\overline{B}_{0_x}$ is critical
for the dynamo. Figure~\ref{fig_c1Re}a) (left)
shows that $\overline{B}_{0_x}$ gains energy from the
nonlinear term ${A}_{01_x}>0$, which is the product of the MRI
correlation of $\mathbf{u}_1$ and $\mathbf{b}_1$, and loses energy via
${D}_{0_x}$ and ${A}_{02^+_x}<0$, so that ${A}_{01_x}\simeq
\vert{D}_{0_x}+{A}_{02^+_x}\vert$. ${A}_{02^+_x}$ transfers energy
to smaller scales (where it is dissipated), and can again
be interpreted as a nonlinear diffusion of $\overline{B}_{0_x}$.

To understand how energy is injected into the MRI
wave and transferred to $\Bo$, we now consider the energy budget
for $\mathbf{b}_1$,
\begin{equation}
\label{b1budget}
{\mathbf{\Omega}}_1+{\mathbf{I}}_1+{\mathbf{A}}_1+{\mathbf{D}}_1=\left<
\mathbf{b}_1\circ({\partial\, \mathbf{b}_1}/{\partial t})\right>
\simeq\bf{0}, \quad \text{where}
\end{equation}
\begin{equation}
{\mathbf{\Omega}}_1=-S\langle\overline{b}_{1_y}\overline{b}_{1_x}\rangle\,\mathbf{e}_y, \quad \quad
{\mathbf{I}}_1=  {\mathbf{I}}_{1{L}}+ {\mathbf{I}}_{1{NL}},
\end{equation}
\begin{equation}
 \label{MRIactivemode}
 {\mathbf{D}}_1= -\eta\,\langle (k_{x}(t)^2+k_{y_0}^2)\,\mathbf{b}_1\circ \mathbf{b}_1 \rangle, \quad \quad
 {\mathbf{A}}_1= {\mathbf{A}}_{10}+{\mathbf{A}}_{12^+},
 \end{equation}
${\mathbf{I}}_{1{L}}=\langle \mathbf{b}_1 \circ
(\Bo\cdot \mathbf{\nabla}\mathbf{u}_1)\rangle$ and
${\mathbf{I}}_{1{NL}}$ is a nonlinear induction
term; ${\mathbf{A}}_{10}=-{\mathbf{A}}_{01}=-\langle
\mathbf{b}_1 \circ  ({\mathbf{u}_1\cdot \mathbf{\nabla}\,\Bo)} \rangle$
is the energy exchanged through advection between the MRI wave and
$\Bo$, and ${\mathbf{A}}_{12^+}$ accounts for a similar
exchange with smaller scales (Eq.~(\ref{b1budget}) is almost
zero because the wave carries a negligible amount of energy at both
$t=0$ and $t=T/2$). Figure~\ref{fig_c1Re}a-b) (right)
shows that the $x$ and $y$ components of the MRI
perturbation $\mathbf{b}_{1}$ are fed by induction (respectively by
$I_{1_x}$ and ${\Omega}_{1_y}$, with $I_{{1NL}_x}\ll {I}_{{1L}_x}$).
As expected, some of the energy injected via the MRI is transferred
back nonlinearly to $\overline{B}_{0_x}$ through
${A}_{{10}_x}=-{A}_{01_x}$, and some of it is lost through laminar
dissipation ${D}_{1_x}$. The rest ${A}_{12^+_x}<0$ is transferred to
smaller scales 

and can be regarded as a nonlinear diffusion of MRI-unstable perturbations. 
Using a similar analysis, we checked that $j\geq 2$ perturbations are
mostly excited via nonlinear interactions, and not by the MRI, for all
cases studied here. Summing Eqs.~(\ref{B0budget}) and (\ref{b1budget})
to eliminate $A_{01_x}$, we obtain
\begin{equation}
\label{eq_budget}
{I}_{0_x}+{I}_{1_x}\simeq\vert{D}_{0_x}+{D}_{1_x}\vert+\vert{D}_{t_x}\vert~,
 \end{equation}
which translates that the energy injected via the MRI as
${I}_{1_x}$ must balance the total of ``laminar'' ohmic dissipation
$\vert{D}_{0_x}+{D}_{1_x}\vert$ and nonlinear dissipation
$\vert{D}_{t_x}\vert\equiv\vert{A}_{12^+_x}+{A}_{02^+_x}\vert$
for the dynamo to be sustained (${I}_{0_x}\ll{I}_{1_x}$ in all cases
studied here). 
 \begin{figure}
\centering
\includegraphics[width=1.\columnwidth]{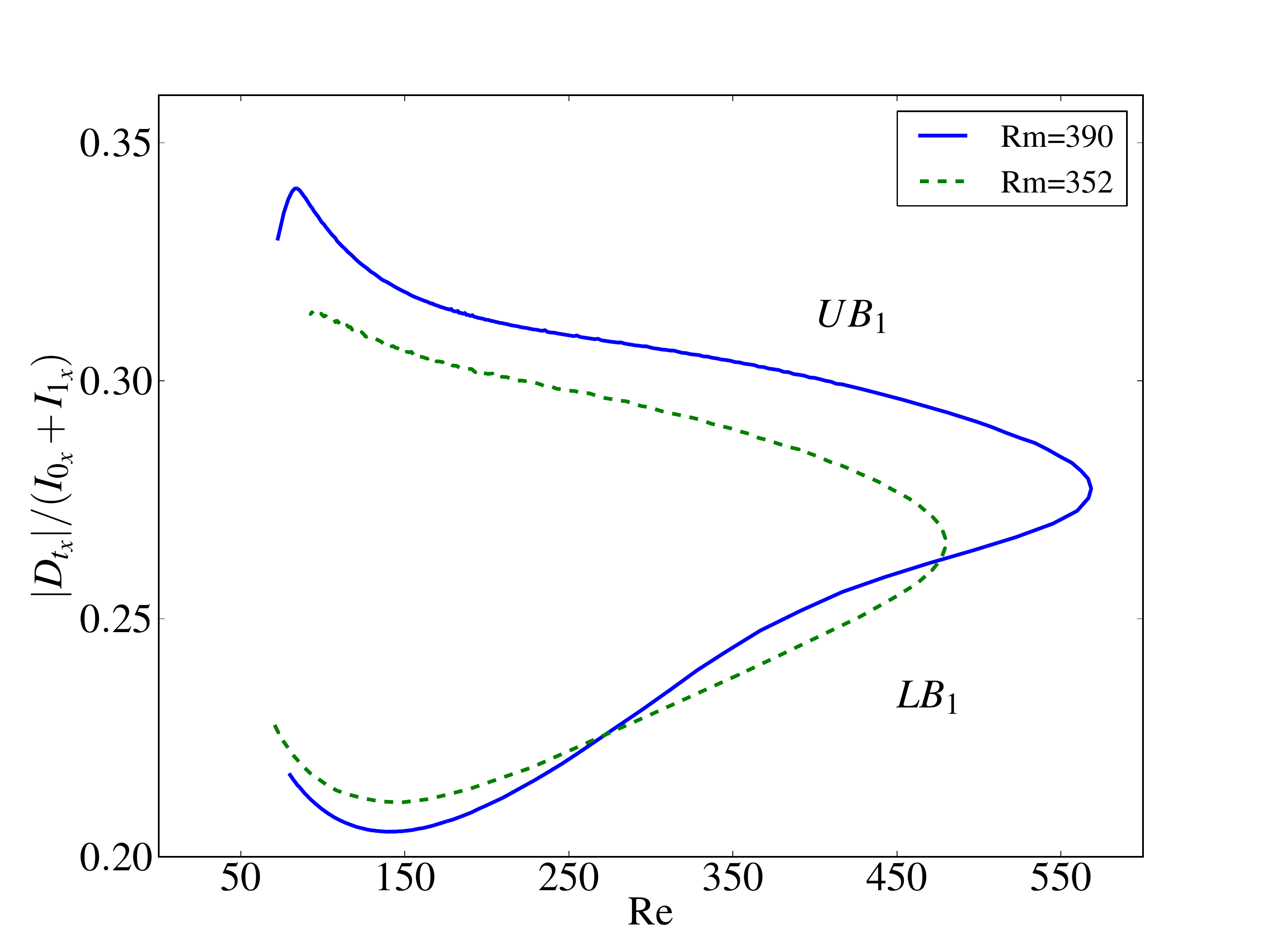}
 \caption{Ratio between
   nonlinear dissipation ${D}_{t_x}$ and injected energy
   ${I}_{0_x}+{I}_{1_x}$ for $LB_1$ and $UB_1$ and two different
   $\text{Rm}$.}
\label{fig_c1Re_ratio}
 \end{figure}

To understand how Eq.~(\ref{eq_budget}) is satisfied in different
regimes, we show in Fig.~\ref{fig_c1Re_ratio} the ratio
$\vert{D}_{t_x}\vert/({I}_{0_x}+{I}_{1_x})$ for $LB_1$ and
$UB_1$ as a function of $\text{Re}$, for two values of
$\text{Rm}$. This ratio is always significantly larger for $UB_1$ than
for $LB_1$, which is consistent with the standard picture
of upper branches being more nonlinear than lower
branches. The amplitude of $\Bo$ is larger on $UB_1$
(Fig.~\ref{fig_frontiere} (inset)), which results in a stronger
MRI driving  (the  MRI is always in the regime
$k_{y_0}\overline{B}_{0_y}<\Omega$ here, so the stronger the field,
the larger the growth rate). The most important observation, however, 
is the significant increase of
$\vert{D}_{t_x}\vert/({I}_{0_x}+{I}_{1_x})$ with $\text{Re}$ for
the saddle solution $LB_1$, which shows that a larger fraction of the
energy injected in $\Bo$ and $\mathbf{b}_1$ is lost through nonlinear
dissipation ${D}_{t_x}={A}_{12^+_x}+{A}_{02^+_x}$ as $\text{Re}$
increases. Our interpretation is that this relative
enhancement of ``turbulent'' magnetic diffusion is tied to the
facilitated excitation of velocity fluctuations at large
$\text{Re}$. 

\section{Disappearance of the dynamo}
\begin{figure}
\centering
\includegraphics[width=1.\columnwidth]{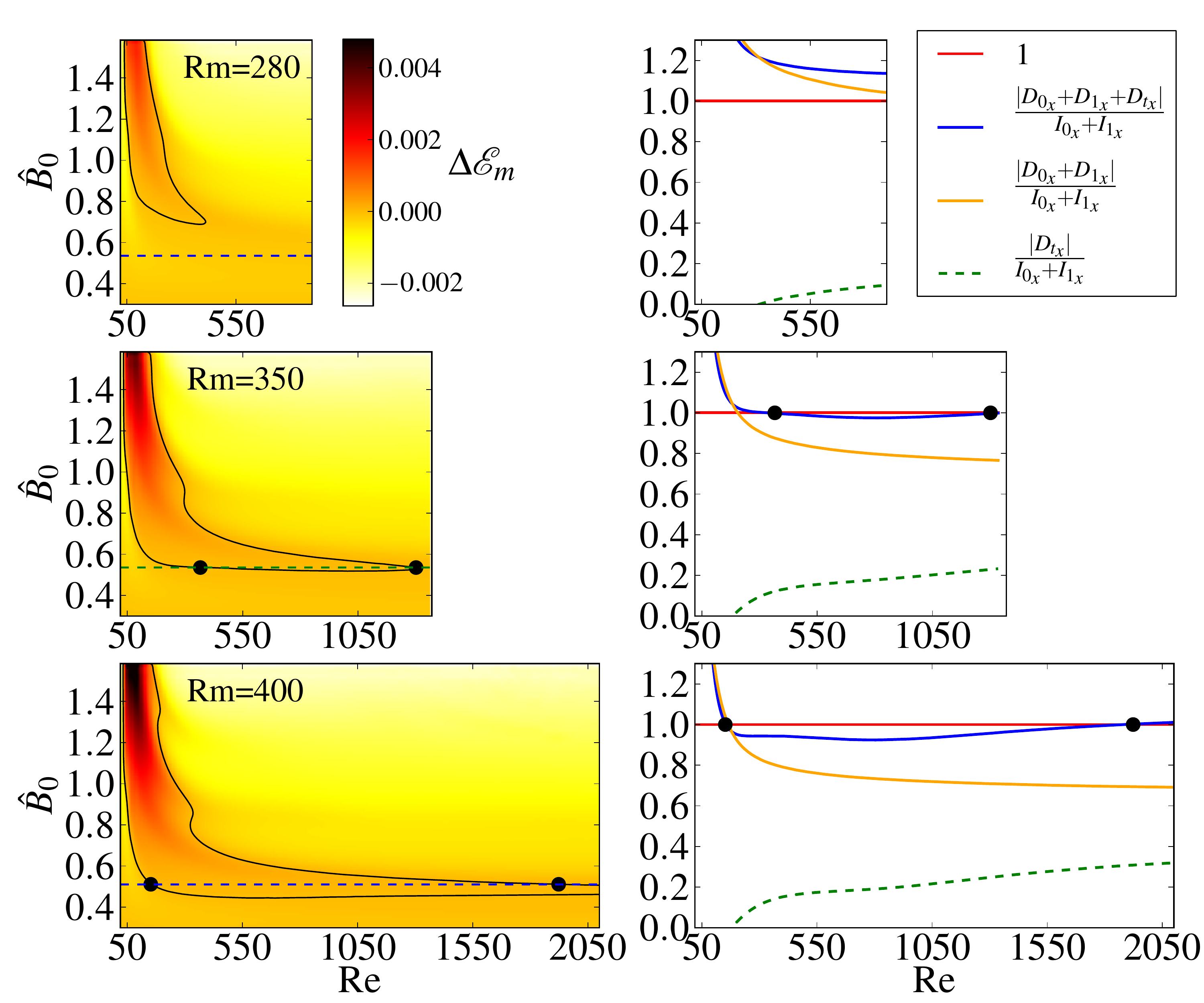}
 \caption{Energy budget of simulations integrated over $St=L_y/L_x$
   for three values of $\text{Rm}$. Left: net
   energy gain $\Delta\mathcal{E}_m$ as a function of $\text{Re}$ and
   $\hat{B}_0$. $\Delta\mathcal{E}_m=0$ isolines are shown in
   black, bullets mark the points $\Delta\mathcal{E}_m=0$
   for $\hat{B}_0\simeq 0.52$. Right: normalized
   injection and dissipation terms in Eq.~(\ref{eq_budget}), as a
   function of $\text{Re}$, for the same $\hat{B}_0$.}
\label{figReBy}
 \end{figure}

The previous results provide qualitative clues to understand the conditions
of excitation of the dynamo. We note that the amplitude of $\Bo$ on
$LB_1$ seems to asymptote to a constant at large $\text{Re}$ and fixed
$\text{Rm}$ (Fig.~\ref{fig_frontiere} (inset)), and therefore so
should the MRI growth rate for this branch. This suggests that the MRI
may not be able to sustain $\Bo$, $\mathbf{b}_1$ and therefore the
dynamo against the total effective magnetic diffusion beyond some
critical $\text{Re}$, as observed in Figs.~\ref{fig_frontiere} and
\ref{fig_c1Re_ratio}.
To investigate more precisely how the energy balance of
Eq.~(\ref{eq_budget}) may be broken, we prepared a family of initial
conditions resembling $LB_1$ at $t=0$, consisting of an axisymmetric field 
$\Bo=\hat{B}_0\,(0.04\,\mathbf{e}_x+\mathbf{e}_y)$ parametrized by
$\hat{B}_0$, plus non-axisymmetric perturbations in the form
of a given packet of shearing waves ($\vert k_y\vert=k_{y_0}$,
$k_x(t=0)=0$ and multiple $k_z$) with weak but random amplitudes  
($\overline{B}_{0_x}$/$\overline{B}_{0_y}=0.04$ is representative of
$LB_1$ and ensures that $\Omega_{0_y}$ is of the order of $D_{0_y}$). 
This set of initial conditions was then integrated by DNS during
half a cycle period typical of a reversal of $\Bo$, for a range of
$\text{Re}$ and $\hat{B}_0$. The results were used to construct a map of the net energy
$\Delta\mathcal{E}_m={I}_{0_x}+{I}_{1_x}-\vert{D}_{0_x}+{D}_{1_x}+{D}_{t_x}\vert$
gained or lost by the active magnetic modes during the reversal, as a
function of $\text{Re}$ and $\hat{B}_0$, for several
$\text{Rm}$ (Fig.~\ref{figReBy} (left)).
The isolines $\Delta\mathcal{E}_m=0$ are reminiscent of the
continuation curves of $SN_1$ (Fig.~\ref{fig_frontiere} (inset)).
The range of $\text{Re}$ in which the system gains more energy from
the MRI than it dissipates $(\Delta \mathcal{E}_m>0)$ widens
significantly at larger $\text{Rm}$. 

Figure~\ref{figReBy} (right) shows plots of the different
terms in Eq.~(\ref{eq_budget}) normalized by $({I}_{0_x}+{I}_{1_x})$ 
as a function of $\text{Re}$, for $\hat{B}_0\simeq 0.52$. 
At $\text{Rm}=280$, the system loses energy for all $\text{Re}$
after a reversal. At $\text{Rm}=350$, there is a range of
$\text{Re}$ in which more energy is pumped in by the MRI than
dissipated. This range widens at even larger $\text{Rm}=400$.
The transition from a sustained to a decaying regime at
large $\text{Re}$ occurs because $\vert
D_{0_x}+D_{1_x}\vert/({I}_{0_x}+{I}_{1_x})$ tends to a constant 
at large $\text{Re}$, while $\vert{D}_{t_x}\vert/({I}_{0_x}+{I}_{1_x})$
increases slowly. The reason why the dynamo can be sustained at larger
$\text{Re}$ as $\text{Rm}$ increases is that the MRI growth rate is not
asymptotic in $\text{Rm}$ in the transitional $300-500$ $\text{Rm}$
range. Laminar dissipation is reduced relative to energy injection as
$\text{Rm}$ increases, which partially offsets the increase in
``turbulent'' diffusion at large $\text{Re}$. Equivalently,
we may conclude that this increase at large $\text{Re}$ requires
 to go to larger $\text{Rm}$ to recover the dynamo, as observed in
 Figs.~\ref{map_turb_ARm} and \ref{fig_frontiere} and reported by
 \cite{fromang07b}.

\section{Discussion}
Why is the MRI dynamo in Keplerian flow harder to excite
at low $\text{Pm}$ ? Using a simple numerical setup, we have
found that weakly nonlinear "turbulent" diffusion (in a qualitative,
not strictly mean-field theoretical sense) of large-scale  magnetic
modes makes it increasingly difficult to sustain the dynamo at
moderate $\text{Rm}$ as $\text{Re}$ increases. The significant advective
transfers of magnetic energy to small scales reported by
\cite{fromang07b} in smaller aspect ratio simulations at
large $\text{Re}$ corroborate this conclusion.
A subtle point is that the velocity fluctuations behind turbulent
magnetic diffusion in this subcritical problem are not externally
prescribed but are indirectly transiently excited by the MRI.

Turbulent diffusion has also been measured in turbulent flows of low
$\text{Pm}$ liquid metals \citep{frick10,forest12}, in which it is
strongly suspected of raising (kinematic) dynamo thresholds
\citep{miralles13}. Given the very generic nature of this effect, we
therefore argue that it could be an important determinant of MRI
dynamo excitation in low $\text{Pm}$ rotating shear flows, such as
occur in parts of some accretion disks (some of
which also have low $\text{Rm}$). Besides, the fact that it
also affects MRI-active modes suggests that it may
be linked to the drop in angular momentum transport
reported in net-flux (imposed field) MRI simulations 
at (moderately) low $\text{Pm}$. 

More work is clearly required to connect these results to the full
diversity of simulated and astrophysical regimes, most importantly
the numerically challenging limit $\text{Re}\gg \text{Rm}\gg
1$, and to study to which extent the conclusions pertain to
numerical configurations with different aspect ratio. {A similar preliminary study in smaller aspect ratio boxes} {suggests that the same qualitative conclusions apply in this case. These results will be presented in a future paper.}  Other nonlinear
effects, some of which may qualitatively relate to the mean-field
theoretical $\alpha$ effect with vertical stratification, are also
probably very important in the disk dynamo problem
\citep{branden95,gressel10,davis10,kapyla11,oishi11,
  simon11,blackman12} and will be worthwhile of investigation along
the same lines.

\begin{acknowledgements}
This research was supported by the University Paul Sabatier of Toulouse under an AO3 grant, by the Midi-Pyr\'en\'ees region and by the French National Program for Stellar Physics (PNPS). Numerical calculations were carried out on the CALMIP platform (CICT, University of Toulouse), whose assistance is gratefully acknowledged.
\end{acknowledgements}


\bibliographystyle{aa}
\bibliography{refs}

\end{document}